\documentclass{aa} 

\usepackage{graphicx}  
\usepackage{txfonts}  
          
\begin{document}  
  
\title{New phase diagrams for dense carbon-oxygen mixtures and white
       dwarf evolution}  
  
\author{Leandro G. Althaus\inst{1,2},  
        Enrique Garc\'{\i}a--Berro\inst{3,4},
        Jordi Isern\inst{4,5},
        Alejandro H. C\'orsico\inst{1,2},
        \and
        Marcelo~M.~Miller~Bertolami\inst{1,2}}
\institute{Facultad de Ciencias Astron\'omicas y Geof\'{\i}sicas, 
           Universidad Nacional de La Plata, 
           Paseo del Bosque s/n, 1900 La Plata, Argentina\
           \and
           Member of CONICET, Argentina\\
           \email{althaus,acorsico,mmiller@fcaglp.unlp.edu.ar}     
           \and
           Departament de F\'\i sica Aplicada,
           Universitat Polit\`ecnica de Catalunya,
           c/Esteve Terrades 5, 
           08860 Castelldefels, Spain\\
           \email{garcia@fa.upc.edu}
           \and       
           Institute for Space  Studies of Catalonia,
           c/Gran Capit\`a 2--4, Edif. Nexus 104,   
           08034  Barcelona,  Spain\\
           \and
           Institut de Ci\`encies de l'Espai (CSIC), 
           Facultat de Ci\`encies, Campus UAB, 
           Torre C5-parell, 
           08193 Bellaterra, Spain\\
           \email{isern@ieec.cat}}

\date{\today}

\abstract{Cool  white  dwarfs  are  reliable and  independent  stellar
          chronometers.   The   most    common   white   dwarfs   have
          carbon-oxygen dense  cores.  Consequently, the  cooling ages
          of very cool white  dwarfs sensitively depend on the adopted
          phase  diagram of  the carbon-oxygen binary mixture.}   
         {A  new   phase  diagram  of   dense  carbon-oxygen  mixtures
          appropriate  for  white dwarf  interiors  has been  recently
          obtained  using direct  molecular  dynamics simulations.  In
          this  paper,  we  explore  the consequences  of  this  phase
          diagram in the evolution of cool white dwarfs.}
         {To do  this we employ  a detailed stellar  evolutionary code
          and  accurate initial  white  dwarf configurations,  derived
          from  the full  evolution of  progenitor stars.  We  use two
          different phase  diagrams, that  of Horowitz et  al. (2010),
          which  presents  an  azeotrope,  and the  phase  diagram  of
          Segretain \& Chabrier (1993), which is of the spindle form.}
         {We computed  the evolution  of 0.593 and  $0.878\, M_{\sun}$
          white dwarf models during  the crystallization phase, and we
          found  that  the  energy  released  by  carbon-oxygen  phase
          separation is smaller when the new phase diagram of Horowitz
          et al.   (2010) is used.   This translates into  time delays
          that are  on average  a factor $\sim  2$ smaller  than those
          obtained  when the  phase diagram  of Segretain  \& Chabrier
          (1993) is employed.}
         {Our  results  have important  implications  for white  dwarf
          cosmochronology, because the cooling  ages of very old white
          dwarfs are  different for the  two phase diagrams.  This may
          have a  noticeable impact on the age  determinations of very
          old   globular   clusters,  for   which   the  white   dwarf
          color-magnitude  diagram  provides  an  independent  way  of
          estimating their age.}

\keywords{stars:  evolution  ---  stars:  interiors ---  stars:  white
          dwarfs}
  
\titlerunning{New  carbon-oxygen   phase  diagrams  and   white  dwarf
              evolution}
  
\authorrunning{L. G. Althaus et al.}  

\maketitle 

 
\section{Introduction}  
\label{intro}  

White  dwarf stars  constitute the  most common  end-point  of stellar
evolution --- see, for instance,  Althaus et al.  (2010a) for a recent
review --- and as such are valuable in constraining several properties
of a wide  variety of stellar populations including  globular and open
clusters (Von Hippel  \& Gilmore 2000; Hansen et  al.  2007; Winget et
al.  2009; Garc\'\i  a--Berro et al. 2010). Additionally,  they can be
used to place constraints on exotic elementary particles (Isern et al.
1992; C\'orsico  et al.   2001; Isern et  al. 2008) or  on alternative
theories of gravitation (Garc\'\i  a--Berro et al.  1995; Benvenuto et
al. 2004; Garc\'\i a--Berro et  al. 2011). This is possible because we
have  a relatively precise  knowledge of  the main  physical processes
responsible  for their  evolution, although  some  uncertainties still
persist for  key aspects  of their constitutive  physics.  One  of the
processes   that   is  still   subject   to   some  uncertainties   is
crystallization.   As early  recognized  (Van Horn  1968), cool  white
dwarfs  are expected  to crystallize  as  a result  of strong  Coulomb
interactions in their very dense interior.  Crystallization results in
two additional energy sources that  slow down the cooling process. The
first source  is latent heat, while  the second one is  the release of
gravitational energy  resulting from the changes  in the carbon-oxygen
profile due to  the different chemical compositions of  the liquid and
solid  phases (Garc\'\i  a--Berro  et al.   1988a, 1988b).   Generally
speaking, the  solid formed upon  cyrstallization is richer  in oxygen
than the liquid.  As the oxygen-rich solid core grows at the center of
the white dwarf, the lighter  carbon-rich liquid mantle left behind is
efficiently redistributed  by Rayleigh-Taylor instabilities  (Isern et
al.   1997).  This  process  releases gravitational  energy, and  this
additional  energy source  has a  substantial impact  in  the computed
cooling times of cool white dwarfs (Segretain et al.  1994; Salaris et
al.  1997; Montgomery et al., 1999;  Salaris et al. 2000; Isern et al.
2000; Renedo et al.  2010).

Recently,  Winget et  al.  (2009)  have used  theoretical fits  to the
observed  white  dwarf luminosity  function  of  the globular  cluster
NGC~6397 to provide evidence  for the occurrence of crystallization in
deep  interiors of  white  dwarfs,  and to  place  constraints on  the
crystallization temperature of the carbon-oxygen mixture.  Thus, it is
foreseable  that in  the near  future deep  photometry of  the cooling
sequence of degenerate stars of nearby globular clusters will allow us
to check the accuracy of  the theoretical cooling sequences, and hence
will allow us to study the  constitutive physics of matter at the high
densities of cool white dwarfs.

Since the pionering works  of Stevenson (1980) and Mochkovitch (1983),
large theoretical efforts have been paid to study the phase diagram of
carbon-oxygen   mixtures.   In   these  early   efforts   the  adopted
carbon-oxygen phase  diagram had a  deep eutectic. This resulted  in a
high  enhancement  of  the   oxygen  abundance  of  the  solid  phase.
Consequently,  the computed  delays in  the cooling  ages  were rather
large.   The  calculations   of  Stevenson  (1980)  were  subsequently
improved by  Ichimaru et al. (1988),  who obtained a  phase diagram of
the azeotropic  form, and  Barrat et al.   (1988), who found  that the
phase diagram was of the spindle form. Later, Ogata et al. (1993) used
Monte  Carlo simulations  and the  hypernetted-chain  approximation to
compute  the phase diagram  of the  carbon-oxygen binary  mixture, and
obtained a  phase diagram of the spindle  form. Almost simultaneously,
Segretain  \& Chabrier  (1993) used  a density-functional  approach to
obtain the phase  diagram for arbitrary binary mixtures  as a function
of $Z_1/Z_2$,  being $Z_1$  and $Z_2$ the  charge of the  two chemical
species.  In the case of carbon-oxygen mixtures, they obtained a phase
diagram of the  spindle type. Since in the case of  a phase diagram of
the spindle  form the  solid phase is  less oxygen-enriched,  the time
delays computed  using this  type of phase  diagrams turned out  to be
smaller  than  those previously  computed  (Segretain  et al.   1994).
Since  then, the  phase diagram  of Segretain  \& Chabrier  (1993) has
remained  a  ``de facto''  standard  over  the  years.  However,  very
recently the phase diagram of dense carbon-oxygen mixtures appropriate
for white dwarf star interiors has been re-examined by Horowitz et al.
(2010).   This work was motivated by the unexpected finding of Winget et al. 
(2009) that the crystallization  temperature of white dwarfs in the globular cluster
NGC~6397  was close to that  for pure carbon. Horowitz et al. (2010)  used  
an  approach  completely   different  of  those
previously  employed.    Specifically,  they  used   direct  two-phase
molecular dynamics simulations for the solid and liquid phases.  Their
results are  in rather good agreement  with those of  Medin \& Cumming
(2010),  and predict  crystallization temperatures  considerably lower
than those  obtained by  Segretain \& Chabrier  (1993).  In particular,
Horowitz et al. (2010) found  the crystallization  temperature of 
carbon-oxygen mixtures with equal mass fractions to be close to that of pure carbon, thus offering
a possible explanation for the puzzling result of Winget et al. (2009). 
 Additionally,
Horowitz  et al.   (2010) found  that the  shape of  the carbon-oxygen
phase diagram is of the azeotropic  form, and not of the spindle type,
as  previously  thought.   This  may   have  a  large  effect  on  the
evolutionary ages of cool white dwarfs.

In  this  paper  we  explore  the implications  for  the  evolutionary
properties of white  dwarfs of the new phase  diagram of carbon-oxygen
mixtures computed by Horowitz et  al.  (2010).  To this end, we employ
a detailed stellar evolutionary  code and initial accurate white dwarf
structures derived  from the full evolution of  progenitor stars.  The
paper  is  organized  as  follows.  In  Sect.~\ref{tools}  we  briefly
describe  the phase  diagram  of  Horowitz et  al.   (2010).  We  also
comment  on  the  main   characteristics  of  our  evolutionary  code.
Particularly we also elaborate on  the treatment of the energy sources
resulting from crystallization, and the evolutionary sequences used in
our study.  Sect.~\ref{results} is devoted to explore the consequences
of the new  phase diagram for the evolutionary  times of white dwarfs.
Finally, in Sect.~4 we summarize  the main results of our calculations
and we draw our conclusions.


\section{Numerical tools} 
\label{tools}

\subsection{Carbon-oxygen phase diagrams}

\begin{figure}  
\centering  
\includegraphics[clip,width=250pt]{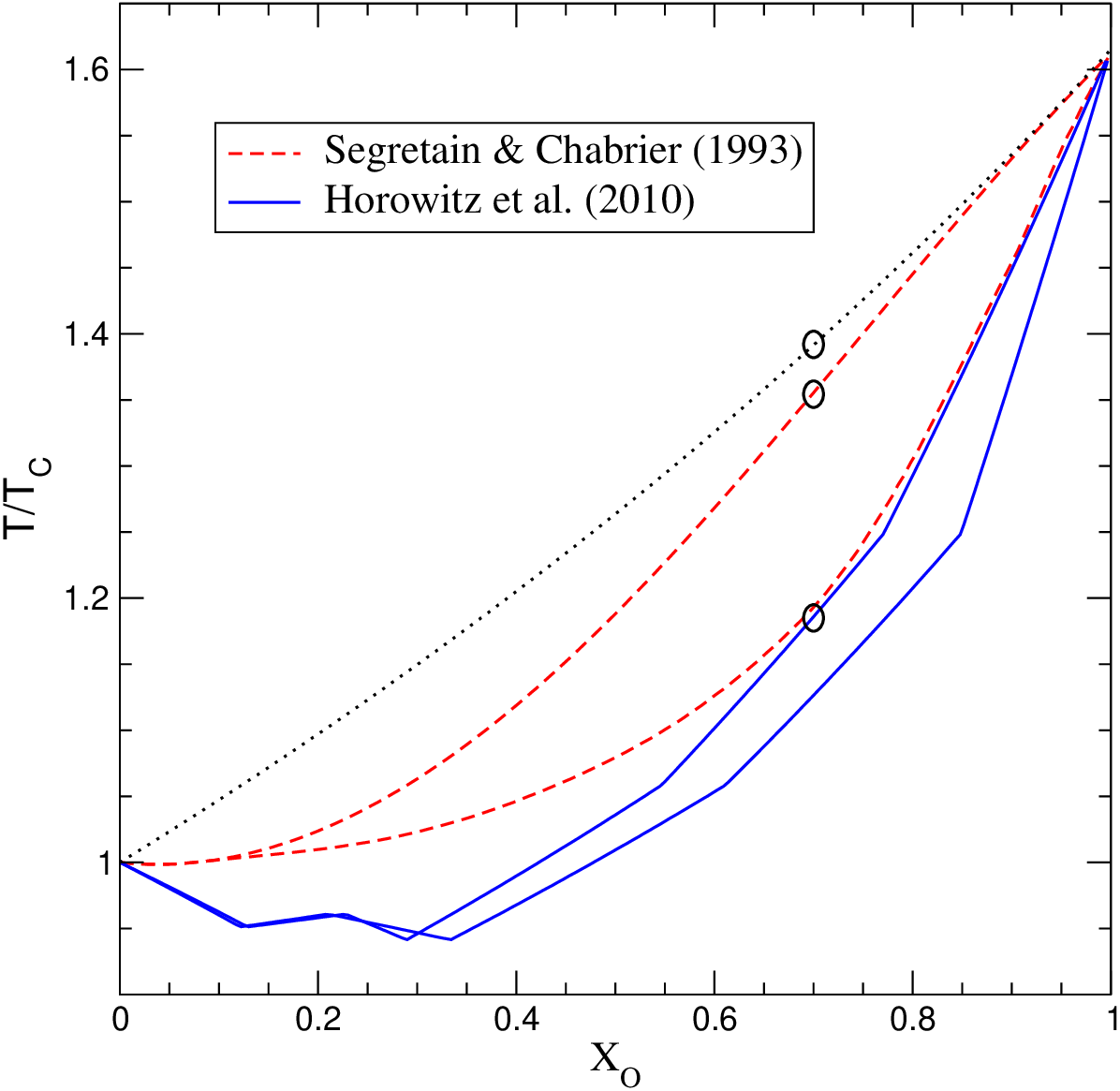}  
\caption{Carbon-oxygen  phase   diagrams  used  in   our  evolutionary
         calculations.    The  crystallization   temperature   of  the
         carbon-oxygen binary mixture  in terms of the crystallization
         temperature  of pure  carbon is  shown as  a function  of the
         oxygen abundance  by mass.  Dashed  red lines and  solid blue
         lines  correspond,  respectively, to  the  phase diagrams  of
         Segretain  \& Chabrier  (1993) and  Horowitz et  al.  (2010).
         For each  diagram, the upper curve  gives the crystallization
         temperature for a given oxygen abundance of the liquid, while
         the lower curve provides  the equilibrium oxygen abundance of
         the solid  at this temperature.  The  dotted line corresponds
         to  the case  in which  no phase  separation occurs,  and the
         mixture is  treated as the  average of the  chemical species.
         Finally, the  circles denote the  crystallization temperature
         of the carbon-oxygen  mixture for the case of  an oxygen mass
         abundance of  0.7, a typical value found  in the evolutionary
         calculations of white dwarf progenitors.}
\label{diagrama}  
\end{figure}  

Horowitz et al.  (2010) have  recently determined the phase diagram of
dense  carbon-oxygen   mixtures  appropriate  for   white  dwarf  star
interiors.  They  argue that their results, based  on direct two-phase
molecular dynamics  simulations, are less affected by  small errors in
the free  energy difference between  the liquid and solid  phases than
previous  studies.  Moreover,  they predict  that  the crystallization
temperature of the carbon-oxygen  binary mixture is considerably lower
than that  resulting from the  phase diagram of Segretain  \& Chabrier
(1993).  This can be  seen Fig.~\ref{diagrama}, in which several phase
diagrams for the carbon-oxygen  mixture are displayed.  In particular,
in  this figure we  show both  the crystallization  temperature (upper
curves) as  a function of the  oxygen abundance by mass  of the liquid
phase and the equilibrium abundances in the solid phase (lower curves)
at  this  temperature.  The  crystallization  temperature  $T$ of  the
carbon-oxygen  mixture is  expressed in  terms of  the crystallization
temperature of  pure carbon  $(T_{\rm C})$.  Crystallization  for pure
carbon occurs  when the Coulomb coupling  parameter ($\Gamma$) reaches
the value $\Gamma_{\rm crys}= 178.4$ (Horowitz et al. 2010), where the
Coulomb coupling parameter is defined as:

\begin{equation}
\Gamma  \equiv \frac{ \langle Z^{5/3} \rangle  e^2}{a_{\rm e} k_{\rm B} T},
\label{gamma}
\end{equation}

\noindent  being $a_{\rm  e}$ the  interelectronic  distance, $\langle
Z^{5/3} \rangle  $ an  average (by number)  over the ion  charges, and
$k_{\rm B}$ Boltzmann's  constant. The rest of the  symbols have their
usual meaning. In particular, the crystallization temperature for pure
carbon composition results

\begin{equation}
T_{\rm C}  = 6^{5/3}\ 2.275 \times 10^5 \frac{(\varrho/2)^{1/3}}{\Gamma_{\rm crys}}.
\label{tc}
\end{equation}

The  shape  of  the  phase  diagram for  a  carbon-oxygen  mixture  as
calculated   by   Horowitz   et    al.    (2010)   is   displayed   in
Fig.~\ref{diagrama} using solid blue lines.  In addition, we also show
(dashed  red lines) the  carbon-oxygen phase  diagram of  Segretain \&
Chabrier (1993).  Finally,  the crystallization temperature when phase
separation is  not taken into account,  and the mixture  is treated as
the average  of the  chemical species, is  represented as  well (black
dotted line).  It is worth highlighting some important features of the
phase diagrams illustrated in Fig.  \ref{diagrama}.  To begin with, we
note  that the phase  diagram of  Horowitz et  al.  (2010)  predicts a
crystallization temperature of the carbon-oxygen mixture substantially
lower  than  that predicted  by  the  phase  diagram of  Segretain  \&
Chabrier (1993), which  in turn is also much  lower than that obtained
in the  case in which  phase separation is disregarded.   This implies
that  for  a  white  dwarf  of  a given  mass  and  core  composition,
crystallization will  set in at smaller stellar  luminosities when the
phase  diagram of  Horowitz  et  al. (2010)  is  adopted.  The  second
relevant point is that the width  of the phase diagram of Segretain \&
Chabrier (1993)  is considerably larger  than that of Horowitz  et al.
(2010).   Hence,  the oxygen  enhancement  in  the  solid phase  (with
respect to  the composition of the  fluid phase from  which it formed)
will be  substantially smaller when  the phase diagram of  Horowitz et
al.  (2010) is adopted.   These two differences have opposite effects.
On the  one hand, the  gravitational energy released  by carbon-oxygen
phase  separation will  be  smaller in  the  case in  which the  phase
diagram of  Horowitz et  al.  (2010) is  adopted.  On the  other, this
energy is released  at smaller luminosities.  The impact  of these two
effects   on  the   delays   introduced  by   phase  separation   upon
crystallization   can  only   be  reliably   assessed  using   a  full
evolutionary code.  Finally, there is  as well another effect which is
also  worth noting.   The phase  diagram  of Horowitz  et al.   (2010)
presents an azeotrope at $X_{\rm  O}\sim 0.3$, while that of Segretain
\& Chabrier (1993)  is approximately of the spindle  form.  This means
that  when the  fluid phase  reaches the  azeotropic  composition, the
solid  phase has the  same composition  of the  liquid.  Consequently,
phase separation no longer occurs, and the subsequent evolution during
the  crystallization phase  is only  driven by  the release  of latent
heat.  Nevertheless, if the abundance of oxygen in the outer layers is
smaller than the  azeotropic one, the solid that  forms is oxygen poor
as compared with the liquid  and raises until it melts, leaving behind
a liquid that gradually  aproaches to the azeotropic composition. When
this happens a solid with the azeotropic composition forms. Therefore,
in this  case, the  process of separation  continues until  the entire
white dwarf core has frozen.

\subsection{Evolutionary code}

All  the calculations  reported here  have  been done  using the  {\tt
LPCODE} stellar evolutionary  code.  This code has been  used to study
different  problems related to  the formation  and evolution  of white
dwarfs ---  see Althaus  et al.  (2010b),  Renedo et al.   (2010), and
references therein for details.   The only difference with respect to
previous  evolutionary calculations  of  cooling white  dwarfs is  the
treatment  of  crystallization.    For  the  results  presented  here,
crystallization  sets in  according to  the phase  diagram considered.
That  is, when  no  phase separation  is  assumed the  crystallization
temperature   of    the   carbon-oxygen   core    is   obtained   from
Eq.~(\ref{gamma}), and imposing $\Gamma=180$, while in all other cases
this temperature is obtained from the corresponding phase diagram.

The  energy sources  associated to  the crystallization  of  the white
dwarf  core comprise  the  release of  latent  heat and  gravitational
energy associated with changes in the carbon-oxygen profile induced by
crystallization.  In  {\tt LPCODE}, the inclusion of  these two energy
sources is done self-consistently and  locally coupled to the full set
of  equations  of  stellar  evolution.   That is,  the  structure  and
evolution of  white dwarfs is  computed with the  changing composition
profile  and with  the luminosity  equation appropriately  modified to
account for  both the local  contribution of energy released  from the
core chemical  redistribution and  latent heat.  At  each evolutionary
timestep,  the  crystallization  temperature  and the  change  of  the
chemical profile  resulting from  phase separation are  computed using
the  appropriate phase  diagram.  In  particular,  the carbon-enhanced
convectively-unstable liquid  layers overlying the  crystallizing core
are  assumed  to be  instantaneously  mixed,  a reasonable  assumption
considering the long evolutionary timescales of white dwarfs (Isern et
al.   1997).  After computing  the chemical  compositions of  both the
solid  and liquid phases  the net  energy released  in the  process is
assessed  as  in  Isern  et   al.   (1997;  2000).   The  latent  heat
contribution is taken to be $0.77  k_{\rm B}T$ per ion (Salaris et al.
2000).  Both  energy contributions are  distributed over a  small mass
range around the crystallization front.  We mention that the magnitude
of both  energy sources  are calculated at  each iteration  during the
convergence  of the  model  --- see  Althaus  et al.  (2010c) for  the
numerical details.

The  input  physics of  the  code include  the  equation  of state  of
Segretain et al. (1994) for the high-density regime --- which accounts
for  all the  important contributions  for both  the liquid  and solid
phases (Althaus et al.  2007) --- complemented with an updated version
of  the  equation of  state  of Magni  \&  Mazzitelli  (1979) for  the
low-density regime. Radiative opacities are those of OPAL (Iglesias \&
Rogers   1996),  including   carbon-  and   oxygen-rich  compositions,
complemented  with  the  low-temperature  opacities  of  Alexander  \&
Ferguson (1994), whilst conductive opacities are taken from Cassisi et
al.  (2007).  During  the white dwarf regime, the  metal mass fraction
$Z$  in the  envelope is  not  assumed to  be fixed.   Instead, it  is
specified  consistently   according  to  the   prediction  of  element
diffusion.   To account  for this,  we considered  radiative opacities
tables   from  OPAL  for   arbitrary  metallicities.    For  effective
temperatures  less than  10,000~K, outer  boundary conditions  for the
evolving models are given  by detailed non-gray model atmospheres that
incorporate  non-ideal  effects  in  the  gas equation  of  state  and
chemical equilibrium (based  on the occupation probability formalism),
radiative and  convective transport (mixing length  theory) of energy,
collision-induced  absorption  from   molecules,  and  the  Ly$\alpha$
quasi-molecular  opacity  (Rohrmann  et  al.   2011).   This  provides
detailed and accurate outer boundary conditions which are required for
a proper treatment of the evolutionary behavior of cool white dwarfs.

\subsection{Evolutionary sequences}

Instead  of exploring  the evolution  of white  dwarfs  with different
arbitrary  chemical profiles for  the core,  we focus  on evolutionary
sequences  that   are  provided   by  detailed  calculations   of  the
evolutionary history of  progenitor stars.  Our aim is  to explore the
impact of the new phase diagram of Horowitz et al.  (2010) on existing
grids  of  white  dwarf  evolutionary  sequences  that  incorporate  a
realistic chemical profile in the stellar interior.  Specifically, the
white  dwarf initial configurations  considered in  this investigation
are  those obtained  from the  full evolution  of progenitor  stars we
computed in previous  studies (Renedo et al. 2010).   In those studies
progenitor stars were evolved from the zero age main sequence, through
the  thermally-pulsing and  mass-loss phases  on the  asymptotic giant
branch (AGB), to the white dwarf cooling phase.  Extra-mixing episodes
beyond  each  formal  convective  boundary were  taken  into  account,
particularly during the core helium  burning stage, but not during the
evolutionary stages corresponding  to the thermally-pulsing AGB phase.
Moreover, the outer chemical profiles of our white dwarf sequences are
the result of  element diffusion processes that lead  to the formation
of pure hydrogen envelopes --- see Althaus et al. (2010b) for details.

\begin{figure}  
\centering  
\includegraphics[clip,width=250pt]{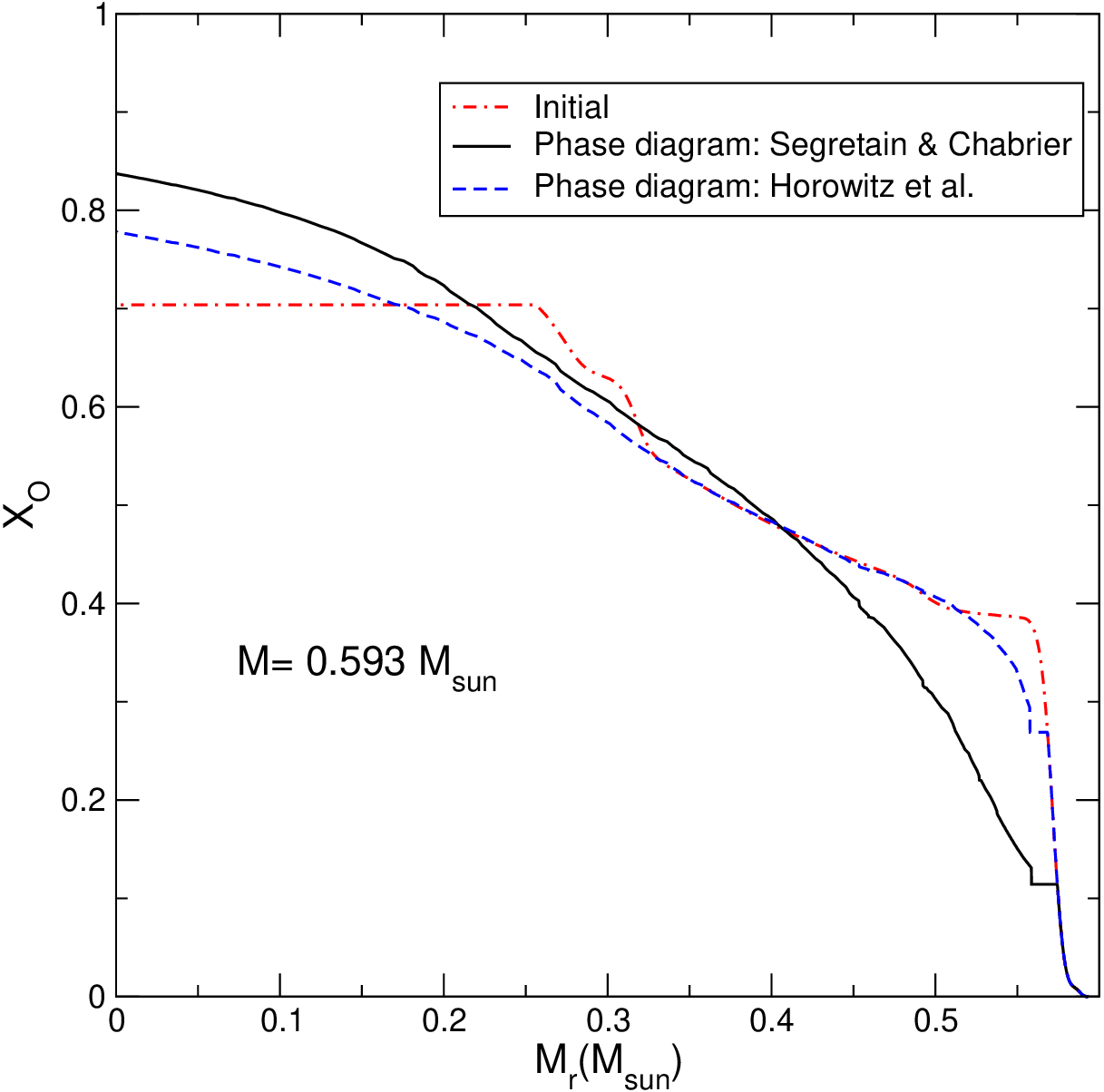}  
\caption{Inner  oxygen  distribution   (abundance  by  mass)  for  the
         $0.593\,   M_{\sun}$   white   dwarf   model   at   different
         evolutionary stages.   The (red) dot-dashed  line corresponds
         to   the  oxygen  distribution   before  the   occurrence  of
         crystallization.  The  final oxygen profile when  most of the
         white  dwarf has  crystallized is  also shown  for  the phase
         diagram of Horowitz et al.  (2010) --- (blue) dashed line ---
         and for  the phase diagram of Segretain  \& Chabier (1993) ---
         (black) solid line.}
\label{perfil_059}  
\end{figure}  

For  the  sake of  conciseness,  we  concentrate  on two  white  dwarf
sequences,  with masses 0.593  and $0.878\,  M_{\sun}$, which  are the
result of the evolution of  1.75 and $5.0\, M_{\sun}$ progenitors with
metallicity $Z=0.01$.   The total mass of hydrogen  in their envelopes
is  $1.1  \times  10^{-4}$   and  $1.17  \times  10^{-5}\,  M_{\sun}$,
respectively.  For each stellar mass, we have computed the white dwarf
cooling phase  down to very low  luminosities, when most  of the white
dwarf  has  already crystallized.   Each  sequence  has been  computed
considering  the phase diagrams  of Segretain  \& Chabrier  (1993) and
Horowitz  et al.   (2010).  In  the interests  of comparison,  we have
computed  additional evolutionary  sequences considering  the chemical
profiles  obtained  using these  two  phase  diagrams  but adopting  a
crystallization temperature  resulting from imposing  $\Gamma=180$. In
this way the effects  of chemical differentiation upon crystallization
can   be   disentangled   from   those  resulting   from   a   smaller
crystallization temperature.


\section{Evolutionary results}  
\label{results}  

\begin{figure}  
\centering  
\includegraphics[clip,width=250pt]{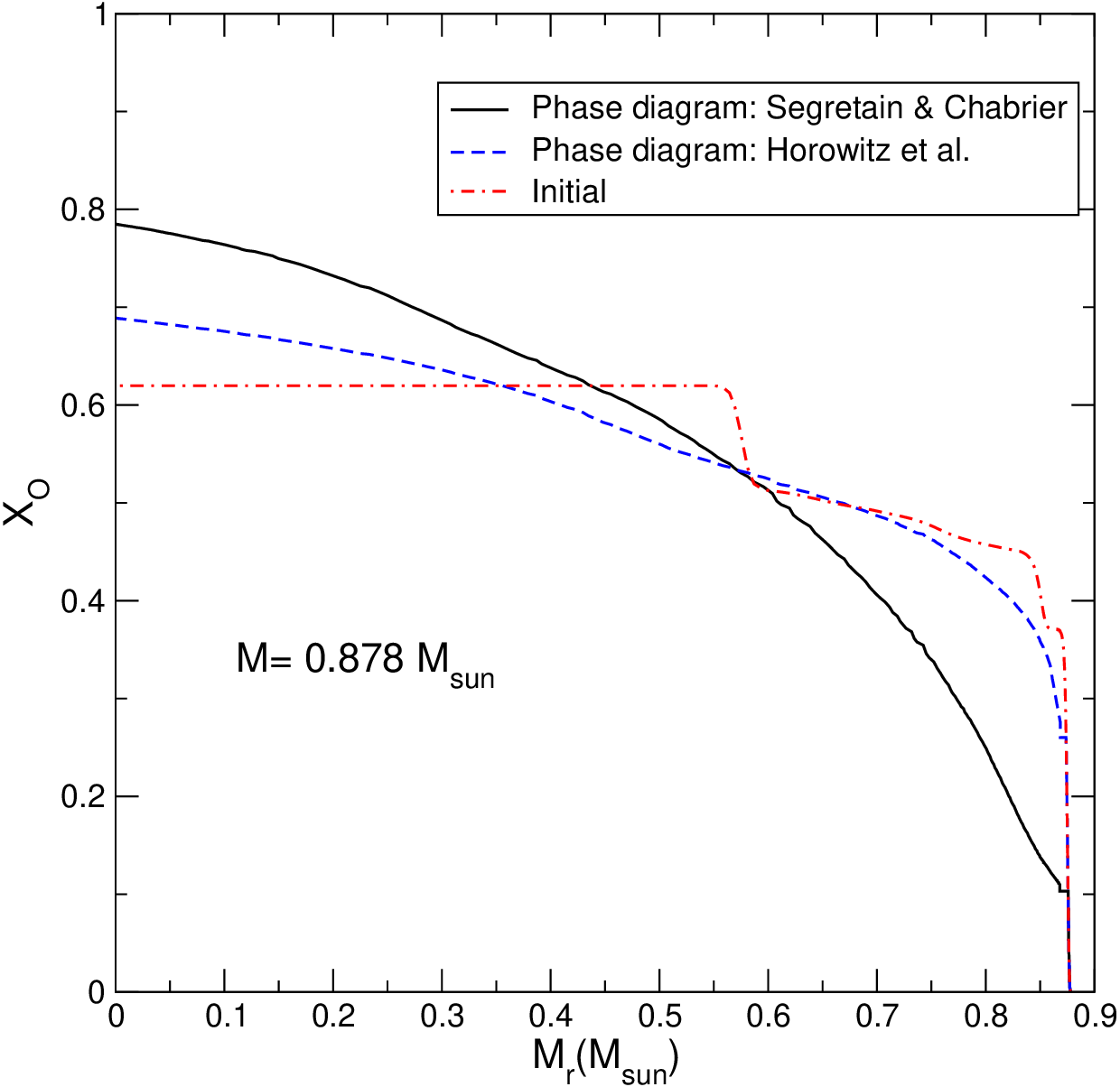}  
\caption{Same  as  Fig.~\ref{perfil_059}  but  for $0.878\,  M_{\sun}$
         white dwarf model sequence.}
\label{perfil_087}  
\end{figure}

We start by examining the impact  of the shape of the phase diagram on
the  oxygen  abundance distribution  expected  in  the  interior of  a
crystallized     white    dwarf.      This    is     illustrated    in
Figs.~\ref{perfil_059} and  \ref{perfil_087}, which display  the inner
oxygen  abundance profile  at  different evolutionary  stages for  the
0.593 and $0.878\, M_{\sun}$ white dwarf models, respectively. In each
figure, the  dot-dashed line shows the oxygen  distribution before the
onset  of  crystallization,  and  the  dashed and  solid  lines  show,
respectively, the oxygen  distribution after crystallization is almost
complete,  when the  phase  diagrams  of Horowitz  et  al. (2010)  and
Segretain  \&  Chabrier  (2003)   are  employed.   As  anticipated  in
Sect.~\ref{tools}, because of the  very different shapes of both phase
diagrams, we expect a distinct  oxygen distribution in the white dwarf
interior by the end of  the crystallization process.  This is apparent
from Figs.~\ref{perfil_059} and \ref{perfil_087}, where it can be seen
that  the  final  oxygen  distributions  predicted by  the  two  phase
diagrams are clearly different.  Note that, for both masses, the phase
diagram of Horowitz et al.  (2010) yields smaller oxygen abundances in
the  central regions,  compared with  those obtained  using  the phase
diagram  of Segretain  \&  Chabrier (1993).   In  particular, for  the
$0.593\,  M_{\sun}$ model, the  crystallization process  increases the
oxygen abundance at the center by  $\sim 10 \%$ when the phase diagram
of Horowitz et al. (2010) is used,  and by $\sim 19 \%$ when the phase
diagram  of Segretain \&  Chabrier (1993)  is adopted.   These figures
turn  out to be  $\sim 11\%$  and $\sim  27\%$, respectively,  for the
$0.878\, M_{\sun}$ model white dwarf.

\begin{figure}  
\centering  
\includegraphics[clip,width=250pt]{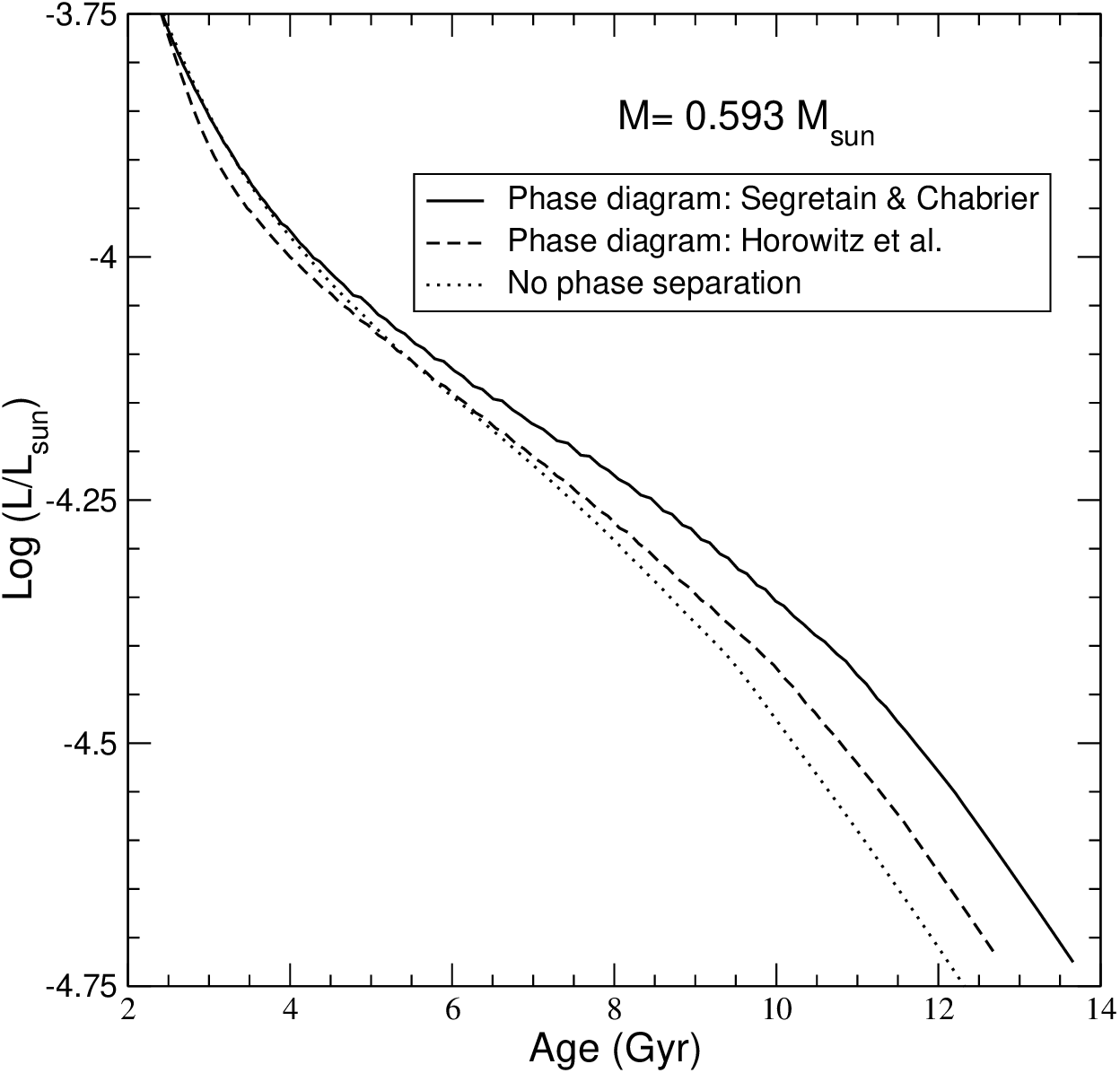}  
\caption{Surface  luminosity  versus age  for  the $0.593\,  M_{\sun}$
         white   dwarf   sequences   undergoing  carbon-oxygen   phase
         separation.  Solid   and  dashed  curves   correpond  to  the
         predictions of  the phase  diagrams of Segretain  \& Chabrier
         (1993)  and  Horowitz  et  al.   (2010),  respectively.   The
         cooling curve in the case that carbon-oxygen phase separation
         is not considered is also shown as a dotted line.}
\label{tcool059}  
\end{figure}

\begin{figure}  
\centering  
\includegraphics[clip,width=250pt]{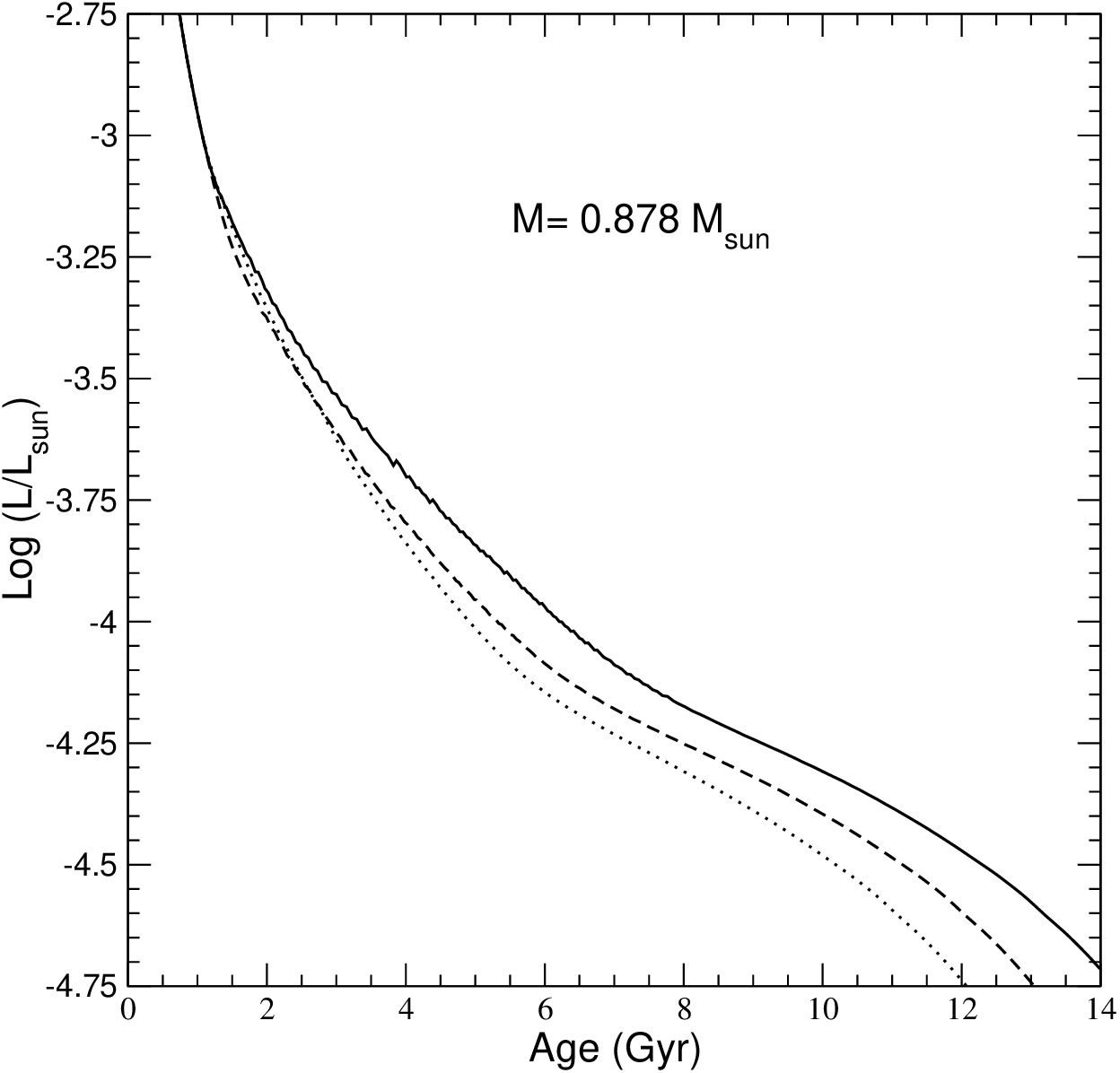}  
\caption{Same as  Fig. \ref{tcool059}  but for the  $0.878\, M_{\sun}$
         white dwarf sequences.}
\label{tcool087}  
\end{figure}

Clearly, the amount of matter redistributed by phase separation during
crystallization for a given stellar  mass is markedly smaller when the
phase diagram of Horowitz et  al.  (2010) is considered.  This results
in a smaller energy  release from carbon-oxygen differentiation. Since
for  the  $0.59\, M_{\sun}$  model  the  initial  oxygen abundance  is
$X_{\rm    O}\sim   0.7$   when    crystallization   sets    in   (see
Fig.~\ref{perfil_059}),  the crystallization  temperature  is not  too
different        for       both       phase        diagrams       (see
Fig.~\ref{diagrama}). Specifically, for  the phase diagram of Horowitz
et al. (2010) the crystallization  temperature is $\sim \, 1.19 T_{\rm
  C}$, whereas  for that of Segretain  et al. (1993) is  $\sim \, 1.35
T_{\rm  C}$.   Consequently,  this   energy  is  released  at  similar
luminosities ---  $\log(L/L_{\sun})\simeq-3.84$ for the  phase diagram
of Horowitz et al.   (2010) and $\log(L/L_{\sun})\simeq-3.70$ for that
 of  Segretain \& Chabrier  (1993) --- and  the impact on  the white
dwarf cooling times is smaller in  the case in which the phase diagram
of Horowitz et  al.  (2010) is adopted.  The same  occurs for the more
massive  white dwarf  (see Fig.~\ref{perfil_087}).   All  this becomes
clear  by  examining  Figs.~\ref{tcool059}  and  \ref{tcool087}  which
depict, respectively, the  relationship between the surface luminosity
and age  for the  0.593 and $0.878\,  M_{\sun}$ white  dwarf sequences
that undergo carbon-oxygen phase separation. In each figure, solid and
dashed curves  correpond to the  predictions of the phase  diagrams of
Segretain   \&  Chabrier   (1993)   and  Horowitz   et  al.    (2010),
respectively.   In  addition,  the  cooling curve  obtained  when  the
carbon-oxygen  phase separation upon  crystallization is  neglected is
shown with  a dotted line  --- for this sequence,  crystallization and
the release of latent heat  are assumed to occur at $\Gamma=180$. Note
that for  both stellar  masses, the phase  diagram of Horowitz  et al.
(2010) results in white dwarf  cooling times that are  smaller
than those  predicted by  the phase diagram  of Segretain  \& Chabrier
(1993).   This  is,  as   mentioned,  a  consequence  of  the  smaller
composition  changes predicted by  the Horowitz  et al.   (2010) phase
diagram.

The dependence  of the age delays induced  by chemical differentiation
on stellar mass  for both phase diagrams is  more difficult to assess,
since they depend  not only on gravity but also  on the temperature at
which  crystallization occurs. Since  gravity is  larger for  the more
massive white dwarf, the energy released by chemical redistribution is
larger  as   well.   However,  for   the  more  massive   white  dwarf
crystallization takes  place at  higher stellar luminosities,  and the
delay in  the cooling times introduced by  chemical differentiation is
smaller in this case --- see, for instance, Salaris et al. (1997).  In
addition,  the magnitude  of  the  age delays  are  influenced by  the
initial chemical  profile of the  white dwarf, which is  different for
each stellar  mass.  This is more  important in the case  of the phase
diagram  of Segretain  \& Chabrier  (1993), for  which the  age delays
strongly depend on the initial  composition (Salaris et al. 2000), but
less  relevant  in  the case  of  the  phase  diagram of  Horowitz  et
al. (2010),  for which the predicted composition  changes are smaller,
see Fig.~\ref{diagrama}.

\begin{figure}  
\centering  
\includegraphics[clip,width=250pt]{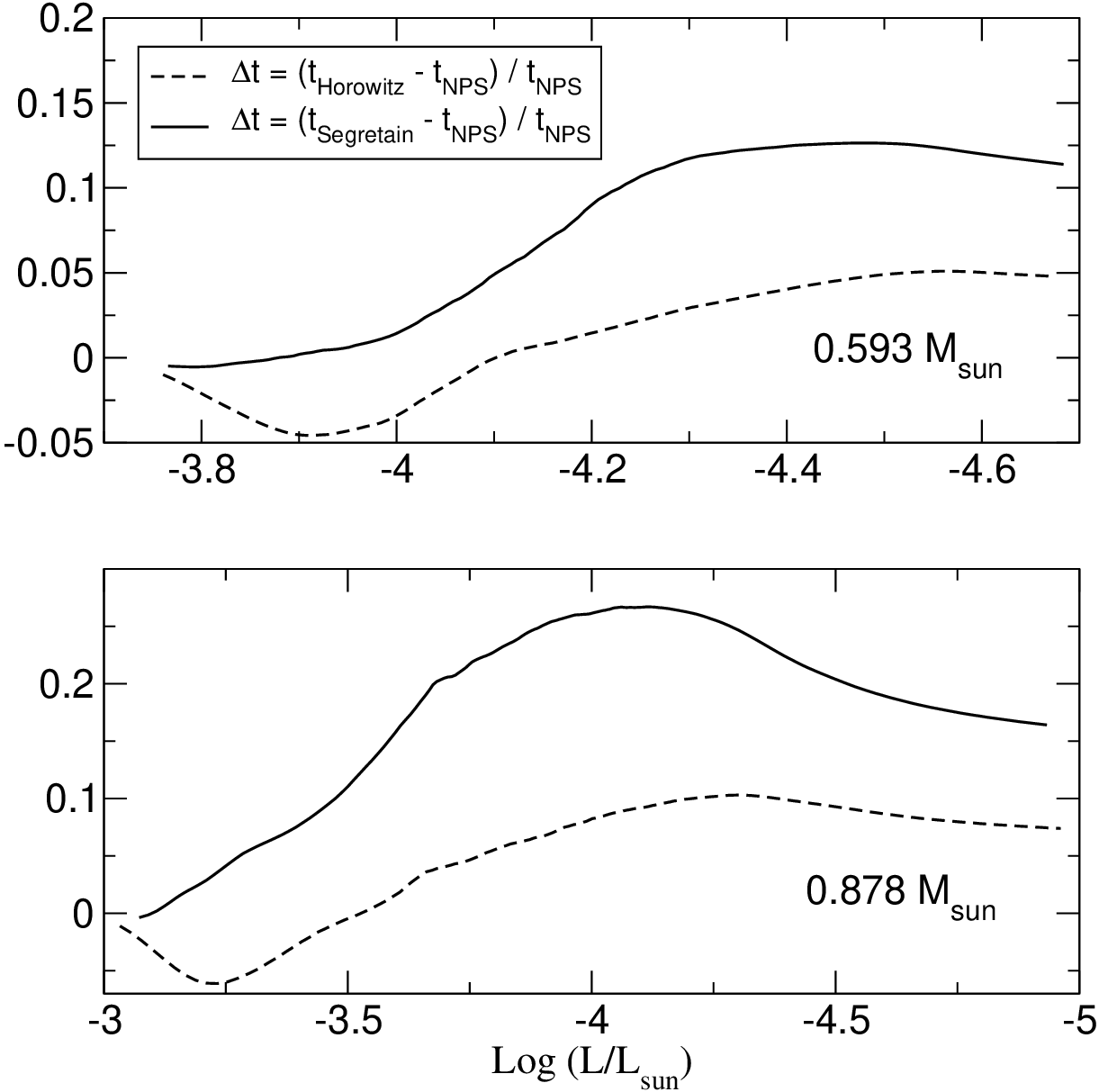}  
\caption{Age delays, in  percentage with respect to the  case in which
         phase  separation  is not  considered,  due to  carbon-oxygen
         phase separation  according to the phase  diagram of Horowitz
         et al.  (2010) and Segretain  \& Chabrier (1993),  dashed and
         solid lines, respectively.  Upper and bottom panel correspond
         to the $0.593$ and  $0.878\, M_{\sun}$ white dwarf sequences,
         respectively.}
\label{dif_sf}  
\end{figure}

The  situation can be  clarified with  the help  of Fig.~\ref{dif_sf},
which illustrates  the age  delays resulting from  carbon-oxygen phase
separation  (in percentage  with respect  to the  case in  which phase
separation  is disregarded)  for  the $0.593$  and $0.878\,  M_{\sun}$
white dwarf  cooling sequences  (top and bottom  panel, respectively),
for both the phase diagram of Horowitz et al.  (2010) --- dashed lines
--- and that  of Segretain  \& Chabrier (1993)  --- solid  lines. Note
that when the phase diagram of Horowitz et al. (2010) is employed, the
age  delays  are negative  at  moderately  high luminosities,  between
$-3.8\ga \log(L/L_{\sun})\ga  -4.1$ for the  $0.593\, M_{\sun}$ model.
This is  simply because the phase  diagram of Horowitz  et al.  (2010)
predicts a lower crystallization  temperature than that obtained using
Eq.~(\ref{gamma}) with $\Gamma=180$.  Hence, crystallization occurs at
lower  stellar luminosities  when  the phase  diagram  of Horowitz  et
al. (2010)  is employed, with  the consequence that  the corresponding
cooling sequence has initially shorter cooling ages.  It is also worth
emphasizing that when the phase  diagram of Horowitz et al.  (2010) is
adopted, the  bulk of the  delay in the  cooling ages of  an otherwise
typical  white  dwarf of  $0.593\,  M_{\sun}$  occurs at  luminosities
ranging  from  $\log(L/L_{\sun})\sim  -4.2$  to  $\log(L/L_{\sun})\sim
-4.5$, whereas for  the phase diagram of Segretain  \& Chabrier (1993)
this  occurs at luminosities  between $\log(L/L_{\sun})\sim  -4.0$ and
$-4.4$, when a sizable part  of the core is already crystallized.  But
the most  relevant aspect that Fig.~\ref{dif_sf} puts  forward is that
when  the  phase  diagram  of  Horowitz et  al.   (2010)  is  adopted,
carbon-oxygen phase  separation is less relevant for  the evolution of
white  dwarfs.    In  particular,   for  the  $0.593\,   M_{\sun}$  at
$\log(L/L_{\sun})\simeq -4.5$,  the age  delay is only  $\sim 0.5$~Gyr
when the phase diagram of Horowitz et al. (2010) is adopted, whilst in
the case  in which that  of Segretain \&  Chabrier (1993) is  used the
delay amounts to $\sim 1.3$~Gyr.

We  also note  that for  the  $0.593\, M_{\sun}$  white dwarf  cooling
sequence, the phase  diagram of Horowitz et al.   (2010) predicts ages
that are up to $\approx  8 \%$ shorter at $\log(L/L_{\sun})=-4.0$ when
compared with the ages derived using the phase diagram of Segretain \&
Chabrier  (1993).   The  differences   are  larger  for  the  $0.878\,
M_{\sun}$ white dwarf sequence, reaching  up to $\approx 17\%$ also at
$\log(L/L_{\sun})=  -4.0$.  In  this case  the differences  are larger
because the smaller initial  oxygen abundance translates into a larger
relative  enrichment of  the solid  phase  when the  phase diagram  of
Segretain \&  Chabrier (1993) is used  --- but not in  the Horowitz et
al.  (2010) phase diagram.  This,  in turn, results in a larger energy
release, and  consequently in larger delays. This  merely reflects the
fact that age delays  resulting from carbon-oxygen separation are less
sensitive to the  initial chemical profile when using  the Horowitz et
al.  (2010) phase diagram.

\begin{figure}  
\centering  
\includegraphics[clip,width=250pt]{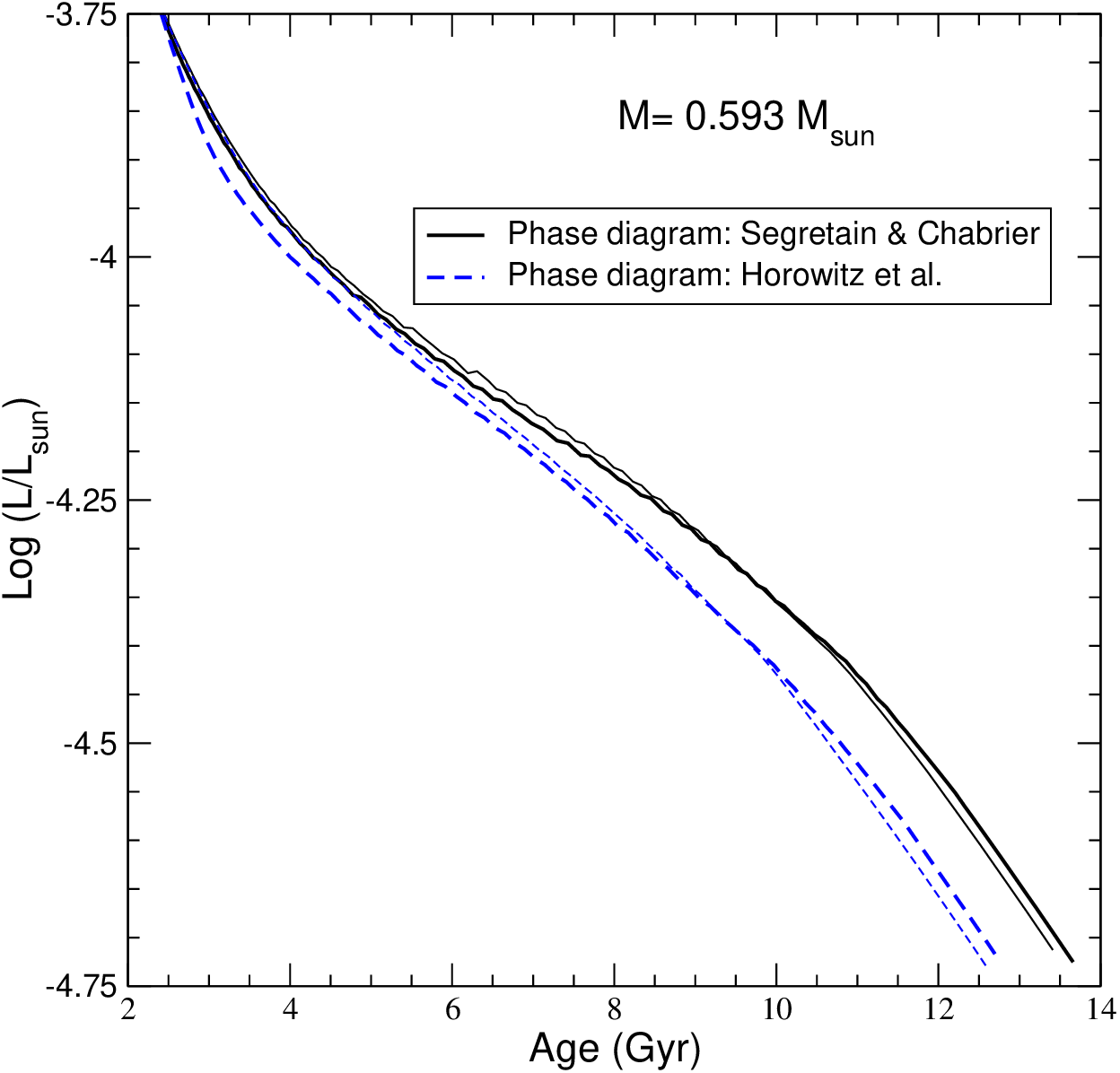}  
\caption{Same as Fig. \ref{tcool059}  but including the cooling curves
         (thin lines)  that result from  imposing that crystallization
         occurs at $\Gamma=180$.}
\label{tcool_varios}  
\end{figure}

To isolate  the effect on the cooling  ages of the shape  of the phase
diagram from  the different crystallization temperature  of both phase
diagrams, in Fig.~\ref{tcool_varios} we compare the cooling curves for
several $0.593\, M_{\sun}$ white  dwarf sequences.  In addition to the
cooling  curves considered  in Fig.~\ref{tcool059},  we also  plot the
cooling   curves   computed  for   both   phase   diagrams  when   the
crystallization  temperature is  given by  the  relation $\Gamma=180$,
rather  than  being  obtained  using  the upper  curve  of  the  phase
diagram. These cooling curves are shown using thin lines.  The results
deserve some comments. As already noted, the phase diagram of Horowitz
et  al.   (2010)  predicts  lower  crystallization  temperatures  (and
luminosities)  than that  of Segretain  \& Chabrier  (1993),  and even
lower than the crystallization temperature when no phase separation is
considered.   Specifically,  for   the  $0.593\,  M_{\sun}$  sequence,
crystallization  starts at  $\log(L/L_{\sun})= -3.70$  when  the phase
diagram   of  Segretain   \&   Chabrier  (1993)   is   used,  and   at
$\log(L/L_{\sun})=  -3.84$ when  that of  Horowitz et  al.   (2010) is
adopted, while when no phase separation is considered, crystallization
occurs  at  $\log(L/L_{\sun})= -3.68$.   This  explains the  initially
larger  cooling   ages  in  the  latter  sequence.    But  at  smaller
luminosities, the trend  is reversed. We thus conclude  that it is the
shape  of  the  phase  diagram   that  is  the  most  relevant  factor
influencing the delays  in the ages of cool white  dwarfs, and not the
specific  value of  the crystallization  temperature.  Finally,  it is
worth mentioning  as well that the  value of $\Gamma$ at  the onset of
crystallization is 186 and 210  for the phase diagrams of Segretain \&
Chabrier (1993) and Horowitz et  al. (2010).  These values turn out to
be  somewhat  higher, 190  and  218,  respectively,  for the  $0.878\,
M_{\sun}$ sequence, because of the initially lower oxygen abundance of
this sequence.

\section{Conclusions}  
\label{conclusion}  

Winget et al. (2009) suggested   that the crystallization  temperature of 
carbon-oxygen white dwarf cores in the globular cluster
NGC~6397 is close to that  of pure carbon. This unexpected result 
prompted Horowitz et al. (2010) to determine a  new phase diagram
for carbon-oxygen mixtures using direct molecular dynamics simulations
for  the   solid  and  liquid  phases.    They  found  crystallization
temperatures considerably  lower than those given by  the most usually
adopted  prescription, which  is obtained  imposing  $\Gamma=180$, and
neglecting carbon-oxygen  phase separation.  In particular,
Horowitz et al. (2010) found that the crystallization temperature for 
carbon-oxygen mixtures with equal mass fractions of carbon and oxygen 
to be close to that of pure carbon, thus offering
an explanation for the puzzling result of Winget et al. (2009).  
They also  found that the
shape of the phase diagram  of binary carbon-oxygen mixtures is of the
azeotropic form,  instead of the  spindle phase diagram  (Segretain \&
Chabrier 1993)  previously employed in the  most accurate calculations
of cooling  white dwarfs (Renedo et  al. 2010) available  so far.  The
core  feature  of  this  paper  has  been  precisely  to  explore  the
consequences for white dwarf evolution  of this new phase diagram.  To
this end, we  used the {\tt LPCODE} stellar  evolutionary code, and we
computed several  cooling sequences for  white dwarfs of  masses 0.593
and $0.878\,  M_{\sun}$.  The initial white  dwarf configurations were
extracted from the full and  detailed evolution of progenitor stars we
computed in  previous studies,  which also provided  realistic initial
chemical profiles.
 
The lower crystallization temperature predicted by the Horowitz et al.
(2010) phase diagram means that  the onset of crystallization in white
dwarfs occurs at stellar  luminosities smaller than those predicted by
the  phase diagram  of Segretain  \& Chabrier  (1993).  For  a typical
white  dwarf of mass  $\sim 0.6\,  M_{\sun}$, we  find that  the phase
diagram of  Horowitz et al.  (2010) predicts  crystallization to occur
at   $\log(L/L_{\sun})=  -3.84$,   while  the   luminosity   at  which
crystallization  sets  in  when  the  phase diagram  of  Segretain  \&
Chabrier  (1993) is  used is  $\log(L/L_{\sun})= -3.70$,  and  when no
phase separation  is considered the  white dwarf core  crystallizes at
$\log(L/L_{\sun})= -3.68$.   Additionally, for this  new phase diagram
the   value  of   Coulomb   coupling  parameter   at   the  onset   of
crystallization for  a $0.6\,  M_{\sun}$ white dwarf  is $\Gamma\simeq
210$.

The amount of matter that  is redistributed by phase separation during
crystallization is  notably smaller in  the new phase diagram  than in
previous calculations of this kind.  Hence, we find that carbon-oxygen
phase separation becomes less  relevant for white dwarf evolution when
this phase diagram is adopted.   At the luminosities for which a large
fraction of the white dwarf  mass has crystallized, we find age delays
due to  carbon-oxygen phase  separation that are  on average  a factor
$\sim 2.5$ smaller than the delays obtained using the phase diagram of
Segretain \& Chabrier (1993).   Another interesting feature of the new
phase diagram of  Horowitz et al.  (2010) is  that composition changes
are  less sensitive  to  the  initial chemical  profile  of the  white
dwarf. This is a relevant point,  since it means that the magnitude of
the age delays induced by  carbon-oxygen phase separation will be less
affected  by  current   uncertainties  in  the  initial  carbon-oxygen
composition of white dwarfs.

Our results  have implications for  the age determinations  of stellar
populations using  the white dwarf  cooling sequence, which  should be
investigated in  subsequent works.  This is  particularly relevant for
the well-studied, old, metal-poor globular cluster NGC~6397, which has
been imaged down  to very faint luminosities and  for which a reliable
color-magnitude  diagram and  a white  dwarf luminosity  function have
been derived.   As mentioned, this cluster has  been recently used to  constrain the
properties  of crystallization  in  the deep  interior  of cool  white
dwarfs (Winget  et al.   2009).  In particular,  Winget et  al. (2010)
have  shown  that the  observed  white  dwarf  luminosity function  in
NGC~6397  seems  to  be  consistent with  pure carbon core white
dwarfs crystallizing at  $\Gamma \approx$ 178,    or, 
alternatively, carbon-oxygen core  white  dwarfs
crystallizing  at  $\Gamma$ values larger than 178,  
the theoretical  value  for  a one
component  plasma. This finding is in line  with  the  predicions of  molecular
dynamics  simulations  of  Horowitz  et al.   (2010). These results together
with the ones reported in this paper call for the need of
studying the cooling
sequence of crystallizing white dwarfs in other old
stellar clusters on the basis of Horowitz et al. (2010) phase diagram. 
Work  in this direction is
in progress.


\begin{acknowledgements}  
This   research   was   supported    by   AGAUR,   by   MCINN   grants
AYA2008--04211--C02--01  and  AYA08-1839/ESP,  by  the  ESF  EUROCORES
Program  EuroGENESIS  (MICINN grant  EUI2009-04170),  by the  European
Union   FEDER   funds,  by   AGENCIA:   Programa  de   Modernizaci\'on
Tecnol\'ogica BID 1728/OC-AR, and  by PIP 2008-00940 from CONICET.
\end{acknowledgements}  
  

\end{document}